\documentclass[notitlepage,aps,11pt,amsmath,amsfonts,amssymb,nofootinbib,superscriptaddress]{revtex4-1}
\usepackage{enumerate}
\usepackage{graphicx}

\usepackage[colorlinks=true, linkcolor=red, citecolor=blue]{hyperref}

\newtheorem{def2}{Definition}
\newtheorem{thm}{Theorem}
\newtheorem{prf}{Proof of Theorem}
\begin{document}

\newcommand{\new}[1]{\ensuremath{\blacktriangleright}#1\ensuremath{\blacktriangleleft}}

\title{Hamilton geometry:\\ Phase space geometry from modified dispersion relations}

\author{Leonardo Barcaroli}
\email{leonardo.barcaroli@roma1.infn.it}
\affiliation{Dipartimento di Fisica, Universit\`a "La Sapienza''
and Sez. Roma1 INFN, P.le A. Moro 2, 00185 Roma, Italy}

\author{Lukas K. Brunkhorst}
\email{ lukas.brunkhorst@zarm.uni-bremen.de}
\affiliation{Center of Applied Space Technology and Microgravity (ZARM),
   University of Bremen, Am Fallturm, 28359 Bremen, Germany}

\author{Giulia Gubitosi}
\email{g.gubitosi@imperial.ac.uk}
\affiliation{Theoretical Physics, Blackett Laboratory, Imperial College, London SW7 2AZ, United Kingdom.}

\author{Niccol\'o Loret}
\email{niccolo.loret@roma1.infn.it}
\affiliation{Dipartimento di Fisica, Universit\`a "La Sapienza''
and Sez. Roma1 INFN, P.le A. Moro 2, 00185 Roma, Italy}

\author{Christian Pfeifer}
\email{christian.pfeifer@itp.uni-hannover.de}
\affiliation{Institute for Theoretical Physics, Universit\"at Hannover, Appelstrasse 2, 30167 Hannover, Germany}
\affiliation{Center of Applied Space Technology and Microgravity (ZARM),
	University of Bremen, Am Fallturm, 28359 Bremen, Germany}

\begin{abstract}
We describe the Hamilton geometry of the phase space of particles whose motion is characterised by general dispersion relations.
In this framework  spacetime and  momentum space are naturally curved and intertwined, allowing for a simultaneous description of both spacetime curvature and non-trivial momentum space geometry.
We consider as explicit examples two models for Planck-scale modified dispersion relations,  inspired from  the $q$-de Sitter and $\kappa$-Poincar\'e quantum groups. In the first case we find the expressions for the momentum and position dependent curvature of spacetime and momentum space, while for the second case the manifold is flat and only the momentum space possesses a nonzero, momentum dependent curvature. In contrast, for a dispersion relation that is induced by a spacetime metric, as in General Relativity, the Hamilton geometry yields a flat momentum space and the usual curved spacetime geometry with only position dependent geometric objects.
\end{abstract}

\maketitle

\newpage

\section{Introduction}
%

The possibility that particles' dispersion relations are modified at the Planck scale is one of the most studied scenarios in quantum gravity phenomenology research \cite{AmelinoCamelia:2008qg,Liberati:2013xla, Mattingly:2005re}.
The main reason is that some astrophysical observations \cite{AmelinoCamelia:1997gz, Ellis:2009yx} are reaching a sensitivity level that allows to test the consequences of modified dispersion relations on the time of propagation of particles \cite{Ackermann:2009aa, AmelinoCamelia:2009pg, Amelino-Camelia:2013naa}. Another, more recent, reason for interest lies in the realization that such effects are relevant also in the very early universe and could provide viable alternatives to the inflationary model \cite{Amelino-Camelia:2013gna, Amelino-Camelia:2013wha, Amelino-Camelia:2013tla, Magueijo:2008yk}.
From a theoretical point of view, modifications of relativistic particle kinematics emerge in several approaches to the quantum gravity problem \cite{Freidel:2005me, Freidel:2003sp, Matschull:1997du, AmelinoCamelia:1999pm, KowalskiGlikman:2002jr, Girelli:2006sc}. 

It is now well understood that Planck-scale modifications of dispersion relations can be encoded in nontrivial geometrical properties of momentum space \cite{AmelinoCamelia:2011bm, KowalskiGlikman:2003we, Gubitosi:2013rna, Majid:1994cy, KowalskiGlikman:2002ft, Arzano:2014jua}. When considering Planck-scale-modified kinematics on flat spacetime, one obtains a picture that is somewhat complementary to the one describing departures from flat-spacetime dispersion relations induced by spacetime curvature. In fact, when looking at the modifications of particles' dispersion relation due to curvature of spacetime \footnote{We are interested in a regime where quantum effects can be neglected, so when we talk about particles we are considering classical objects. Even in this case it is still relevant to look at the Planck-scale regime, since we are essentially taking a limit where $\hbar\rightarrow 0$ but the Planck energy stays finite \cite{AmelinoCamelia:2011bm, AmelinoCamelia:2011pe}. Moreover, we look at dispersion relations written in terms of the physical energy-momentum quantities, which are in general different from the conserved charges under space-time translations.}, one will in general introduce some modifications depending on a distance scale which is related to the curvature itself, see more detailed discussion around eq. (\ref{eq:2}) below. On the other hand, modifications of particles' dispersion relations depending on some energy scale (usually assumed to be the Planck energy), will signal that  momentum space is curved, with the curvature radius related to the energy scale entering the dispersion relation.
So in the first case spacetime is assumed as possibly curved base manifold and momenta belong to its cotangent bundle. In the other case it has been demonstrated that one can choose momentum space as the possibly curved base manifold and then coordinates live on its cotangent bundle \cite{AmelinoCamelia:2011pe}. Of course, when neither spacetime nor momentum space curvature are present, one reduces to the standard special-relativistic scenario and either spacetime or momentum space can be chosen as base manifold.

When both spacetime curvature and Planck-scale deformations of momentum space are present, it is expected that the nontrivial geometry of momentum space and spacetime get intertwined \cite{Marciano:2010gq}, so that giving a geometrical description of either of them becomes highly nontrivial.
This is however the case mostly of interest for the purpose of phenomenology, since for both the astrophysical  observations and, even more, in the early universe context, curvature of spacetime is relevant.
Until now only some preliminary work has been done in this direction \cite{Cianfrani:2014fia, AmelinoCamelia:2012it}. Ref. \cite{AmelinoCamelia:2012it} has the merit of  clearly highlighting the phenomenological significance of the interplay between spacetime curvature and non-trivial momentum space effects. However, it does not provide a general geometrical framework for encoding models where both spacetime and momentum space are curved. For example, it does give a receipt for handling cases where spacetime is not maximally symmetric (meaning that, in the limit where the momentum space sector of phase space becomes trivial, spacetime reduces to a manifold whose metric is not maximally symmetric). Ref. \cite{Cianfrani:2014fia} does provide a general action principle to describe motion of particles with non-trivial features both in the momentum space and spacetime sectors. However, invariance under momentum space diffeomorphisms is implemented, which has no clear physical interpretation.

Here we propose to use the framework of Hamiltonian geometry of phase space, i.e. the cotangent bundle of spacetime, that naturally allows for a description where spacetime and momentum space are curved and intertwined. The starting point is a Hamiltonian describing the propagation of free particles and one is also required to fix the symplectic structure. Within this framework diffeomorphism invariance under general spacetime diffeomorphisms is implemented, while there is no such invariance under general momentum space diffeomorphisms. Moreover, relativistic symmetries, when present, are simply the symmetries encoded in the Hamiltonian.

To go more into detail,  let us start by looking at a classical particle on a generally curved background.
At each point, a freely falling particle has the dispersion relation\footnote{Here and in the following we set the speed of light $c=1$. We also assume signature $(+,-,-,-)$ for the metric.}
\begin{equation}\label{eq:1}
E^2=\vec{p}\ {}^2+m^2\,.
\end{equation}
Here $E$ is the energy and  $\vec p$ the spatial momentum an observer associates to the particle, while $m$ is the invariant mass parameter of the particle. With the help of the spacetime metric~$g$, its inverse~$g^{-1}$ and the four-momentum $p$ of the particle, 
the dispersion relation can be written covariantly in terms of the Hamiltonian $H_{g}$
\begin{equation}\label{eq:2}
H_g(x,p)=g^{ab}(x)p_ap_b=m^2\,. 
\end{equation}
The relation between equation (\ref{eq:1}) and equation (\ref{eq:2}) is given by the expansion of the latter in an orthonormal frame of the metric associated to the observer. In this sense the dispersion relation is closely intertwined with the geometry of spacetime, given by the spacetime metric~$g$.



When introducing  modifications of the dispersion relation with non-quadratic terms in energy/momentum the geometry cannot be metric spacetime geometry anymore.
It is however always possible to  interpret any dispersion relation as the level sets of a Hamilton function $H$ on phase space
\begin{equation}\label{eq:5}
H(x,p)=M^{2},
\end{equation}
where $M$ is a mass scale associated to the particle.
The Hamiltonian determines the motion of free test particles via the Hamilton equations of motion, and, as we will demonstrate, the geometry of phase space. For the metric Hamiltonian on the left hand side of equation (\ref{eq:2}), which represents the dispersion relation of general relativity, it turns out that the geometry of phase space can be disentangled into the usual metric spacetime geometry of position space and a trivial, flat, momentum space geometry. For a general Hamiltonian, i.e. a general dispersion relation, this disentanglement will no longer be possible and there will only be an intertwined geometry of  position space and momentum space. 

A somewhat similar approach to the analysis of the geometry of dispersion relations, which has been followed in some previous works \cite{Girelli:2006fw, Gibbons:2007iu, Amelino-Camelia:2014rga}, is to perform a Legendre transform from phase space to position and velocity space to obtain a length measure for curves on configuration space defined by a general Finsler function. Then one obtains the Finsler geometry of spacetime induced by the dispersion relation which is in general an intertwined geometry of the position and velocity space of the particle  trajectories. However, this approach faces basically two drawbacks. On the one hand it is highly non-trivial to perform the Legendre transform for non-metric Hamiltonians explicitly and, on the other hand, the Finsler geometry of spacetime is not well-defined as soon as the Finsler function possesses non-trivial null vectors and does not satisfy certain smoothness conditions \cite{Pfeifer:2011tk, Minguzzi:2014hwa}. Most Finsler functions obtained from suggested modified dispersion relations do not satisfy the required criteria to obtain a well-defined Finslerian spacetime geometry.  As a result of all these difficulties, only modifications of the flat-spacetime dispersion relations have been considered in this framework.

In this work we derive the geometry of phase space from the Hamiltonian which corresponds to a given dispersion relation, omitting the problematic step of going from position and momentum space to position and velocity space and thus circumventing difficulties which appear when going to the dual description in the Finsler geometry approach.
%
%

 Mathematically, the Hamiltonian is a function on the cotangent bundle of a spacetime manifold. The geometry of Hamilton spaces is a geometry of the cotangent bundle of spacetime derived solely from the Hamiltonian, in a similar way as in metric geometry the geometry of a manifold is derived from a metric \cite{Miron}. 
 One important result we will discuss is that the curves on phase space which solve the Hamilton equations of motion become autoparallels of the Hamilton geometry of the cotangent bundle, in physics terminology freely falling, only if the Hamiltonian is homogeneous with respect to the momenta. In case the Hamiltonian is not homogeneous the solutions of Hamilton's equations of motion are dragged away from being autoparallels by a force-like term. Moreover, we will demonstrate that the momentum space (seen as a subspace of phase space) becomes naturally curved in Hamilton geometry as soon as the third derivative of the Hamiltonian with respect to the momenta does not vanish.

As we already mentioned, the most prominent advantage to study modified dispersion relations as Hamiltonians with Hamiltonian geometry is that the framework naturally  incorporates a non-trivial curved geometry of position and momentum space consistently at the same time, a feature that is cumbersome in other approaches that study modified dispersion relations. To demonstrate the features of the general framework we will in particular derive the Hamilton geometry of the cotangent bundle induced by dispersion relations inspired form the $q$-de Sitter and $\kappa$-Poincar\'e quantum groups \cite{Ballesteros:2013fh, Lukierski:1991ff, Ballesteros:2004eu, Lukierski:1991pn, Majid:1994cy, Lukierski:1992dt, Lukierski:1993wxa}. The $\kappa$-Poincar\'e quantum group is one of the most studied models in quantum gravity phenomenology  encoding departures from standard  relativistic kinematics  without spoiling the relativity principle, thanks to modified laws of transformations between inertial observers. The $q$-de Sitter quantum group is somewhat less well-known, but it is indeed very promising as it provides a relativistic generalisation of the de Sitter relativistic group, in the same sense as $\kappa$-Poincar\'e  generalizes the Poincar\'e group. 

We present our results as follows. We begin in section \ref{sec:2HamGeom} with the introduction of Hamilton geometry, the framework with which we seek to analyse general dispersion relations. This geometry, built solely on the basis of a Hamilton function, is based on the definition of Hamilton spaces in section \ref{sec:2HamGeom} and the unique Hamilton non-linear connection of the cotangent bundle (of phase space) which we introduce in \ref{ssec:hconn}. It relates the Hamilton equations of motion to autoparallels of the geometry. Moreover the Hamilton non-linear connection enables us to define the curvature of phase space as well as the curvature of momentum and configuration space in section \ref{ssec:curv}. We discuss the symmetry properties of Hamilton spaces in section \ref{ssec:symm}. Having clarified the mathematical language, we explicitly derive the geometry induced by modified dispersion relations that are perturbations of the metric dispersion relation in section \ref{sec:moddr}. First we study general cubic perturbations of the quadratic metric dispersion relations in section \ref{ssec:mdr3}, then   we analyse the $q$-de Sitter dispersion relation in section \ref{ssec:qds}, showing the $\kappa$-Poincar\'e case as the limit where spacetime curvature vanishes. As mentioned above, in this paper we focus on the classical ($\hbar\rightarrow 0$) regime (where the limit is taken in such a way that the Planck energy stays finite). In section \ref{sec:quantisation} we address some issues concerning the extension of our framework to its quantum version ($\hbar\neq 0$). We conclude and give an outlook in section \ref{sec:4conc}. 

In this paper latin indices $a,b,..i,..$ go from $0$ to $N$, where $N+1$ is the spacetime and momentum space dimensionality.

\section{Hamilton Geometry}\label{sec:2HamGeom}
Hamilton geometry is the geometry of phase space determined solely by a Hamilton function on the phase space of free particles, similarly as the geometry of spacetime is derived from a metric in general relativity. More precisely, the phase space is identified with the cotangent bundle $T^*M$ of a manifold $M$ and is equipped with a smooth Hamilton function $H$. The geometry of the cotangent bundle is derived from this function and its derivatives. In this section we recall these mathematical notions and comment on them from the point of view of what is of interest for the purposes of this work, but leave mathematical technicalities mostly aside. A more technical review on the geometry of the cotangent bundle can be found in Appendix~\ref{app:cotb}. 

First we define Hamilton spaces and introduce the canonical non-linear connection which defines their geometry. Then we prove that for every Hamiltonian that is homogeneous with respect to the momenta the Hamilton equations of motion become the autoparallel equation of the Hamilton geometry, i.e. for those Hamiltonians test particles fall freely in the geometry. This is not true anymore for non-homogeneous Hamiltonians. With the help of the fundamental non-linear connection we can define the phase space curvature and identify the curvature of position and momentum space as parts of phase space. Moreover we discuss the notion of symmetries of Hamilton spaces. 

We basically follow the construction of the geometry of Hamilton spaces from the book \cite{Miron} with slight modifications and generalisations necessary to include a wide range of physically interesting Hamiltonians.


\subsection{Hamilton spaces and Hamilton equations}\label{ssec:21hs}
The cotangent bundle $T^*M$ of an n-dimensional manifold $M$ is itself the 2n-dimensional manifold built as the union of all cotangent spaces of $M$
\begin{equation}
T^*M=\bigcup_{q\in M} T^*_qM\,.
\end{equation}
An element $\Omega\in T^*M$ is a one-form on $M$ and in local coordinates $x$ in a neighborhood $U$ around a point in $M$ we can write $\Omega=p_a dx^a_{|x}\in T^*_{x}M\subset T^*M$. Thus we can label $\Omega$ with coordinates $(x,p)$. This procedure yields local coordinates on $T^*M$ called manifold induced coordinates and have the property that the Poisson bracket between these momentum coordinates and the manifold coordinates is the canonical one
\begin{equation}
	\{x^a, p_b\}=\frac{\partial}{\partial x^q} x^a \frac{\partial}{\partial p_q} p_b-\frac{\partial}{\partial x^q} p_b \frac{\partial}{\partial p_q} x^a=\delta^a_b\,.
\end{equation}
It is worth to point out that this choice for the symplectic structure emerges here as natural consequence of the form of $\Omega$. It is possible to map all that will follow in another symplectic choice, obtaining a far more complicated description of the geometry of the phase space, still obtaining the same kinematics~\cite{AmelinoCamelia:2011cv,Amelino-Camelia:2013uya}.\\
A change of coordinates on the base manifold $x\rightarrow x'(x)$ induces a coordinate change of the manifold induced coordinates on $T^*M$, called manifold induced coordinate transformations, according to the transformation behaviour of one-forms on $M$\footnote{Note that the change of coordinates on the base manifold depends on the base manifold coordinates only.}
\begin{equation}\label{eq:mfcoord}
(x_a,p_b)\rightarrow (\tilde x_a,\tilde p_b)=(\tilde x_a(x),p_q\frac{\partial x^q}{\partial \tilde x^b})\,.
\end{equation} 
Seeing $T^*M$ as manifold we immediately obtain the manifold coordinates induced basis of the tangent and cotangent space, $T_{(x,p)}T^*M$ and $T^*_{(x,p)}T^*M$, of $T^*M$ denoted by $\{\partial_a=\frac{\partial}{\partial x^a}, \bar{\partial}^a=\frac{\partial}{\partial p_a}\}$ and $\{dx^a,dp_a\}$. Further mathematical details on the cotangent bundle, like the behaviour of these bases under coordinate changes of the manifold $M$ and their interpretation from the point of view that $T^*M$ is naturally a fibre bundle can be found in appendix~\ref{app:cotbcoord}. Having clarified the notation we can define Hamilton spaces.

\begin{def2}\textbf{Hamilton space}\\
	A Hamilton space $(M,H)$ is an n-dimensional smooth manifold $M$ equipped with a continuous function $H:T^*M\rightarrow\mathbb{R}$ on its cotangent bundle, the Hamiltonian, that satisfies:
	\begin{itemize}
		\item $H$ is smooth on $T^*M\setminus\{0\}$,
		\item the Hamilton metric $g^H$ of $H$ is non-degenerate, nearly everywhere on $T^*M\setminus\{0\}$
		\begin{equation}\label{eq:hammetric}
		g^{Hab}(x,p)=\frac{1}{2}\frac{\partial}{\partial p_a}\frac{\partial}{\partial p_b}H(x,p)=\frac{1}{2}\bar\partial^a\bar\partial^b H(x,p)\,.
		\end{equation}
	\end{itemize} 
\end{def2}
These are minimal assumptions on Hamilton spaces in order to describe the geometry of the cotangent bundle in terms of Hamiltonian geometry. In contrast to the definition in \cite{Miron} we do not require here that the Hamilton metric has constant rank and is non-degenerate everywhere on $T^*M\setminus\{0\}$. 
We should notice at this point that those metrics should be interpreted as a tool to get the non-linar connections and they are not to be confused with Rainbow metrics~\cite{Magueijo:2002xx} nor with momentum-space metrics~\cite{AmelinoCamelia:2011bm}, already known and widely used in literature.\footnote{Rainbow metrics $g_R^{ab}(x,p)$ and momentum space metrics $\zeta^{ab}(x,p)$ both generate dispersion relations, and so Hamiltonians. For Rainbow metrics the relation is
\begin{equation}
H(x,p)=\mathcal{C}(x,p)=g_R^{ab}(x,p) p_a p_b\,,
\end{equation}
while momentum space metrics, employed in the framework of relative locality define the invariant mass parameter of a particle via
\begin{equation}
m \equiv \int_0^1\sqrt{\zeta^{mn}(x(\tau),p(\tau))\dot{\lambda}_m\dot{\lambda}_n}\,d\tau\, .
\end{equation}
It has been shown that in some cases momentum space metrics $\zeta_{ab}(x,\dot{x})$ can define an invariant (under $\ell$-deformed transformations) spacetime line-element~\cite{Loret:2014prd}.}

One example of Hamiltonians that we can include are homogeneous Hamiltonians of the form \begin{equation}\label{eq:Hn}
H(x,p)=G^{a_1...a_n}(x)p_{a_1}...p_{a_n}
\end{equation}
into the definition of Hamilton spaces. They are straightforward homogeneous generalisations of the metric Hamiltonian, which falls into this class for $n=2$. It is known that such Hamiltonians for $n=4$ describe the propagation of light in general linear electrodynamics \cite{Hehl}, for example in non-dissipative optical media \cite{Schuller:2009hn}, they are the duals to Finsler geometries which may be considered as generalisations of metric spacetime geometry to explain astrophysical observations~\cite{Pfeifer:2011xi} and they describe the geometric optical limit of partial-differential equations \cite{Raetzel:2010je}. However for such Hamiltonians the non-degeneracy requirement is usually not satisfied on all of $T^*M\setminus\{0\}$. Even though we will not discuss those Hamiltonians just mentioned in this article, we desire to include such applications into the general formalism. 

In this work, we use non-homogeneous Hamiltonians to study Planck-scale-deformed dispersion relations. For example, the first-order correction to the standard special-relativistic dispersion relation has the general form:
\begin{equation}
H(x,p,\ell)=p_{0}^{2}-\vec p^{\,2}+\ell Q^{a_{1}a_{2}a_{3}} p_{a_{1}}p_{a_{2}}p_{a_{3}}\,,
 \end{equation}
where $\ell^{-1}$ is the energy/momentum scale and $Q^{a_{1}a_{2}a_{3}} $ a matrix of numerical coefficients.

The Hamiltonian encodes the dynamics of point particles via the Hamilton equations of motion
\begin{equation}\label{eq:heom}
\dot p_a + \partial_aH=0,\quad \dot x^a-\bar\partial^a H=0\,.
\end{equation}
These equations determine the trajectory of a point particle in phase space, i.e. in the cotangent bundle $T^*M$ of the spacetime manifold $M$. They immediately imply that the Hamiltonian is conserved along curves $\gamma=(x(t),p(t))$ in $T^*M$ which are solutions of the equations
\begin{equation}
\dot\gamma(H)=\dot x^a\partial_aH+\dot p_a\bar{\partial}^aH=\bar{\partial}^aH(\dot p_a+\partial_a H)=0\,.
\end{equation}
The second Hamilton equation is just the duality map which connects the cotangent bundle of a manifold with the tangent bundle 
\begin{eqnarray}\label{eq:dual}
\sharp:T^*M\rightarrow TM;\quad (x,p)\mapsto \sharp(x,p)=(x,\bar\partial^a H(x,p))=(x, y(x,p))\,,
\end{eqnarray}
while the first Hamilton equation describes the motion of the system in momentum space. Next we construct a connection on the cotangent bundle such that the first Hamilton equation of motion becomes the autoparallel equation of this connection with source term for a general Hamiltonian. 

\subsection{The Hamilton non-linear connection and its autoparallels}\label{ssec:hconn}
The fundamental object in the description of the intrinsic geometry of a manifold is a connection which defines parallel transport and curvature. In metric geometry there exists a unique torsion-free connection which leaves the metric covariantly constant, namely the Levi-Civita connection. In Hamilton geometry we employ the so-called Hamilton non-linear connection which generalises the Levi-Civita connection to the general cotangent bundle setting. Moreover, the Hamilton non-linear connection enables us to study the geometry of momentum space and position space as subsets of phase space consistently at the same time. Further mathematical details on connections on the cotangent bundle are explained in appendix \ref{app:cotbconn}.

An important difference between the case where the Hamiltonian is homogeneous and the more general case we are interested in is that solutions of the Hamilton equations of motions are autoparallels of the connection only for homogeneous Hamiltonians, while for inhomogeneous Hamiltonians there is a force-like term present which prevents equality to autoparallel motion. In theorem \ref{thm:2} we derive the former, which is known in the literature, from the latter.

The definition of the non-linear connection requires use of  the Poisson bracket of functions $F$ and $G$ on $T^*M$
\begin{equation}
\{F(x,p),G(x,p)\}=\partial_a F \bar{\partial}^aG-\partial_a G \bar{\partial}^aF.
\end{equation}
Then we can display the Hamilton non-linear connection as follows:
\begin{def2}\textbf{The Hamiltonian non-linear connection}\label{def:nonlin}\\
	Let $(M,H)$ be a Hamiltonian geometry. Then
	\begin{equation}\label{eq:nlin0}
	N_{ab}(x,p)=\frac{1}{4}\bigg(\{g^H_{ab},H\}+g^H_{ai}\partial_b\bar\partial^i H+g^H_{bi}\partial_a\bar\partial^i H\bigg)
	\end{equation}
	are called connection coefficients of the Hamilton non-linear connection.
\end{def2}
This connection is called non-linear since it may depend non-linearly on the momenta. In terms of these connection coefficients we can define a covariant derivative on the cotangent bundle for so-called d-tensors. An $(r,s)$-d-tensor field on the cotangent bundle is a tensor field which behaves like an $(r,s)$-tensor field on the manifold, regarding the transformation behaviour and the number of components. The difference to an $(r,s)$-tensor field on the manifold is that the components of the d-tensor field depend on positions and momenta, not only on positions. Let $T^{a_1...a_r}{}_{b_1...b_s}(x,p)$ be the components of a d-tensor field. The components of its dynamical covariant derivative are given by
\begin{eqnarray}\label{eq:dyncov}
\nabla T^{a_1...a_r}{}_{b_1...b_s}&=&\{T^{a_1...a_r}{}_{b_1...b_s},H\}+Q^{a_1}{}_{m}T^{ma_2...a_r}{}_{b_1...b_s}+...+Q^{a_r}{}_{m}T^{a_1...m}{}_{b_1...b_s}\nonumber\\
&-&Q^{m}{}_{b_1}T^{a_1...a_r}{}_{mb_2...b_s}-...-Q^{m}{}_{b_s}T^{a_1...a_r}{}_{b_1...m},
\end{eqnarray}
with $Q^a{}_b=2N_{bq}g^{Hqa}-\partial_b\bar{\partial}^aH$. With help of the dynamical covariant derivative we state:
\begin{thm}\label{thm:1}
	The Hamilton non-linear connection coefficients are the unique connection coefficients which satisfy
	\begin{equation}\label{eq:thm1}
	N_{ab}=N_{ba},\quad \nabla g^H_{ab}=0\,.
	\end{equation}
\end{thm}
The symmetry of the connection coefficients is obvious and related to the compatibility of the connection with the symplectic structure. The covariant derivative condition determines the symmetric part of the $N_{ab}$ simply by expanding the condition using the definition of the dynamical covariant derivative. For a metric Hamiltonian $H_g$, see equation (\ref{eq:2}), the Hamilton connection coefficients are basically the Christoffel symbols of the Levi-Civita connection of the metric with components $g_{ab}(x)$ 
\begin{equation}
N_{ab}[H_g](x,p)=-p_q \Gamma^q{}_{ab}(x)\,.
\end{equation}
The transformation behaviour of the connection coefficients allow us to introduce special bases of the tangent and cotangent spaces of the cotangent bundle, the so-called Berwald or horizontal-vertical bases, which transform like basis vector and covector fields on the base manifold under manifold induced coordinate transformations (\ref{eq:mfcoord})
\begin{equation}
T_{(x,p)}T^*M=\text{span}(\delta_a=\partial_a - N_{ab}\bar{\partial}^b, \bar\partial^a),\ T^*_{(x,p)}T^*M = \text{span}( dx^a, \delta p_a=dp_a+N_{ab}dx^a ) \,.
\end{equation}
The part of $T_{(x,p)}T^*M$ which is spanned by the $\delta_a$ is called the horizontal tangent space and the complement spanned by $\bar{\partial}^a$ the vertical tangent space. For the dual space $T^*_{(x,p)}T^*M$ the part which is spanned by the $dx^a$ is called the horizontal cotangent space and the complement spanned by $\delta p_a$ called the vertical cotangent space.
\begin{itemize}
	\item The \emph{vertical spaces} represent the tangent respectively cotangent spaces of \emph{momentum space},
	\item the \emph{horizontal} spaces represent the tangent respectively cotangent spaces of \emph{spacetime},
\end{itemize}
both as subspaces of the tangent respectively cotangent spaces of phase space. Observe that the $\delta_a$ and $\delta p_a$ do not reduce to $\partial_a$ and $dp_a$ in general metric geometry but to $\delta_a=\partial_a+\Gamma^q{}_{ar}p_q\bar{\partial}^r$ and $\delta p_a=dp_a-p_s\Gamma^{s}_{ar}dx^r$, since the spacetime manifold and the momentum space are seen as complementary subspaces of phase space and not as separated spaces on their own. Further mathematical details of the horizontal-vertical split of the tangent spaces of cotangent bundle and the dynamical covariant derivative are discussed in appendix \ref{app:cotbconn}.

Now there exist special curves $\zeta(t)=(x(t), p(t))$ on the cotangent bundle namely those whose tangent is purely horizontal. The requirement for a purely horizontal tangent is
\begin{equation}
\dot \zeta(t)=\dot x^a \partial_a + \dot p_a \bar{\partial}^a=\dot x^a\delta_a+(\dot p_a+N_{ab}\dot x^b)\bar{\partial}^a\overset{!}{=} \dot x^a\delta_a\,.
\end{equation}
Those curves are called autoparallels of the Hamilton non-linear connection. From this we find the autoparallel equation to be
\begin{equation}
\dot p_a+N_{ab}\dot x^b=0\,.
\end{equation}
Comparing this to the Hamilton equations of motion (\ref{eq:heom}) we find the important result that solutions of the Hamilton equations of motions are in general autoparallels of the Hamilton non-linear connection up to a source term
\begin{equation}\label{eq:hamauto}
0=\dot p_a + \partial_a H = \dot p_a + N_{ab}\bar\partial^bH+\partial_a H- N_{ab}\bar\partial^bH=\dot p_a + N_{ab}\bar\partial^bH+\delta_aH \,.
\end{equation}
The physical interpretation of this result is that for general Hamiltonians the motion of particles cannot be understood as free fall motion in a geometry. There is a force-like term $-\delta_a H$ present which drags particles away from free fall. However in the special case where Hamiltonians are homogeneous of any degree $r$ with respect to the momenta $H(x, \lambda p)=\lambda^r H(x,p)$ the following holds, proven in appendix \ref{app:prfthm2}: \footnote{As mentioned earlier, Planck-scale modified dispersion relations cannot be encoded in homogeneous Hamiltonians.}

\begin{thm}\label{thm:2}
	Let $(M,H)$ be a Hamiltonian manifold with homogeneous Hamiltonian $H$, i.e. $H(x,\lambda p)=\lambda^r H(x,p)$, and let $N_{ab}$ the connection coefficients of the Hamilton non-linear connection. Then
	\begin{equation}\label{eq:thm2}
	\delta_a H=\partial_aH-N_{ab}\bar\partial^b H=0\,.
	\end{equation}
\end{thm}
Thus for any homogeneous Hamiltonian we have recovered the statement that particles following the Hamilton equations of motion fall freely on autoparallels of the Hamilton geometry of phase space. 
This means that all forces acting on a test particle which can be described by one homogeneous Hamiltonian can be absorbed into one phase space geometry in which a test particle is freely falling. In particular this statement holds for all polynomial Hamiltonians displayed in equation (\ref{eq:Hn}), so especially, and not surprisingly, for the metric Hamiltonian which describes a test particle on which only the gravitational force acts in general relativity. There equation (\ref{eq:hamauto}) is equivalent to the usual geodesic equation on the metric spacetime, as will be seen explicitly in section \ref{ssec:mdr3}.

\subsection{The curvature of phase space, spacetime and momentum space}\label{ssec:curv}
The curvature of phase space is the curvature of the Hamilton non-linear connection on the cotangent bundle. It measures the integrability of spacetime, i.e. position space, as a subspace of the cotangent bundle and is defined as the commutator between the horizontal vector fields, see definition \ref{def:curvature} in appendix \ref{app:cotbconn} for a mathematical definition, 
\begin{equation}\label{eq:nlincurv}
[\delta_a,\delta_b]=\big(-\delta_a N_{cb}+\delta_b N_{ca}\big)\bar\partial^c=R_{cab}\bar\partial^c\,.
\end{equation}
In general this curvature depends on all phase space coordinates $(x,p)$. For a metric Hamiltonian $H_g$, see equation (\ref{eq:2}), we find that it reduces basically to the well-known Riemann curvature tensor and is linear in the momenta
\begin{equation}
R_{cab}[H_g](x,p)=p_qR^q{}_{cab}(x)\,.
\end{equation}
In any case this curvature of the cotangent bundle intertwines position and momentum space, even for metric phase space geometry. 

We have seen in the previous section that the non-linear connection yields a split of the directions on the cotangent bundle into horizontal and vertical directions. We can think of the vertical directions as directions along momentum space and the horizontal directions as the complementary directions along position space in phase space. While the non-linear connection defines the geometry for the phase space itself as a whole it is possible to associate linear covariant derivatives to the non-linear connection which respect the horizontal-vertical split of the directions, i.e. the split into directions along momentum space and along spacetime. That means they map horizontal vectors onto horizontal vectors and vertical vectors onto vertical vectors:
\begin{eqnarray}
\nabla_{\delta_a}\delta_b&=&F^c{}_{ab}\delta_c,\quad \nabla_{\delta_a}\bar\partial^b=F^b{}_{ac}\bar\partial^c\\
\nabla_{\bar\partial^a}\delta_b&=&E^{ac}{}_b\delta_c,\quad \nabla_{\bar\partial^a}\bar\partial^b=E^{ab}{}_c\bar\partial^c\,.
\end{eqnarray}
These covariant derivatives are defined through their coefficients $F$ and $E$ and define the geometry of momentum space and spacetime as parts of phase space. Observe that they could not be defined without fixing the non-linear connection in advance since the non-linear connection allows us to identify momentum space and spacetime directions in a covariant way with respect to diffeomorphisms of the manifold, due to its transformation behaviour. 

Among all associated covariant derivatives there is a distinguished one called Cartan-linear covariant derivative $\nabla^{CL}$ defined through the coefficients
\begin{equation}\label{eq:lincoef}
F^a{}_{bc}=\frac{1}{2}g^{Haq}(\delta_bg^H_{cq}+\delta_cg^H_{bq}-\delta_qg^H_{cb})=:\Gamma^{\delta a}{}_{bc},\quad E^{ab}{}_c=-\frac{1}{2}g^H_{rc}\bar\partial^a g^{Hrb}=:C^{ab}{}_c\,.
\end{equation}
This Cartan-linear covariant derivative is unique in the sense that its coefficients are both symmetric, that means torsion-free, and leave the Hamilton metric vertically, along momentum space, and horizontally, along spacetime, covariantly constant
\begin{equation}
\nabla^{CL}_{\delta_a}g^{Hbc}=0,\quad \nabla^{CL}_{\bar\partial^a}g^{Hbc}=0\,.
\end{equation}
We can now consider the purely horizontal component $R^q{}_{abc}(x,p)$ of the curvatures of the Cartan-linear covariant derivative and the purely vertical one $Q_q{}^{abc}(x,p)$. These are the curvatures making parallel transport along spacetime, respectively along momentum space non-trivial:
\begin{eqnarray}
R^{Hq}{}_{abc}(x,p)\delta_q&=&\nabla^{CL}_{\delta_b}\nabla^{CL}_{\delta_c}\delta_a-\nabla^{CL}_{\delta_c}\nabla^{CL}_{\delta_b}\delta_a-\nabla^{CL}_{[\delta_b,\delta_c]}\delta_a\\
&=&\big(\delta_b\Gamma^{\delta q}{}_{ac}-\delta_c\Gamma^{\delta q}{}_{ab} +\Gamma^{\delta q}{}_{bi}\Gamma^{\delta i}{}_{ac}-\Gamma^{\delta q}{}_{ci}\Gamma^{\delta i}{}_{ab}-R_{ibc}C^{qi}{}_{a}\big)\delta_q\,,\label{eq:hcurv}\\
Q_{q}{}^{abc}(x,p)\bar{\partial}^q&=&\nabla^{CL}_{\bar\partial^b}\nabla^{CL}_{\bar\partial^c}\bar\partial^a-\nabla^{CL}_{\bar\partial^c}\nabla^{CL}_{\bar\partial^b}\bar\partial^a\\
&=&\big(\bar\partial^b C^{ac}{}_q-\bar\partial^c C^{ab}{}_q+C^{bi}{}_q C^{ac}{}_i-C^{ci}{}_q C^{ab}{}_i\big)\bar{\partial}^q\label{eq:vcurv}\,.
\end{eqnarray}
Our interpretation of the vertical and horizontal curvature as curvature of momentum and position space is consistent with what we know from metric geometry. For a metric Hamiltonian geometry the curvature of momentum space $Q$ vanishes since the $C^{ab}{}_c[H_g]$ vanish, and, as can be easily calculated, the components of the horizontal curvature tensor become the components of the usual Riemann curvature tensor of metric spacetime geometry. Conversely, for a Hamiltonian that does not depend on spacetime, the horizontal curvature vanishes since the $\Gamma^{\delta a}{}_{bc}$ vanish, but the vertical curvature does not necessarily disappear. For a generic Hamiltonian both curvatures depend on the positions and momenta. Thus Hamiltonian phase space geometry enables us to study phase spaces with curved momentum and position space as well as phase spaces where only one of both spaces is curved or none. This is the fundamental aspect in Hamiltonian phase space geometry which makes it so valuable in the description of modified dispersion relations when both spacetime curvature and Planck-scale modifications are present.

\subsection{Symmetries}\label{ssec:symm}
Symmetries of a metric manifold $(M,g)$ are diffeomorphisms of the manifold which leave the metric invariant. 
In the same spirit we say that a diffeomorphism $\Phi$ of $T^*M$, i.e. of phase space, is a symmetry if it leaves the Hamiltonian invariant:
\begin{equation}
 H(\Phi(x,p))=H(x,p)\,. \label{eq:SymmetryCondition}
\end{equation}
From the infinitesimal action of the diffeomorphism,
%
\begin{equation}
\Phi(x,p)=(\tilde x(x,p),\tilde p(x,p))=(x^a+\epsilon \xi^a(x,p),p_a+\epsilon \bar \xi_a(x,p))+\mathcal{O}(\epsilon^2)\,,\label{eq:diffeo}
\end{equation}
one finds the vector field generating the symmetry transformation by asking that the above definition of symmetry, eq. (\ref{eq:SymmetryCondition}), holds:
\begin{eqnarray}
H(\phi(x,p))&=&H(x^a+\xi^a(x,p),p_a+\bar \xi_a(x,p))\nonumber\\
&=&H(x,p)+\epsilon(\xi^a(x,p)\partial_a H(x,p)+\bar \xi_a(x,p)\bar{\partial}^aH(x,p))+\mathcal{O}(\epsilon^2)\nonumber\\
&=&H(x,p)+\epsilon Z(H)(x,p)+\mathcal{O}(\epsilon^2)=H(x,p)\,.
\end{eqnarray}
The vector field $Z=\xi^a(x,p)\partial_a+\bar \xi_a(x,p)\bar{\partial}^a$ on $T^*M$  is then the generator of the diffeomorphism $\Phi$, which has to satisfy the following condition for $\Phi$ being a symmetry
\begin{equation}\label{eq:symmcon}
Z(H)=0\,.
\end{equation}

\begin{def2}\textbf{Symmetry generators}\\
	Let $(M,H)$ be a Hamilton geometry. A generator of a symmetry of $(M,H)$ is a vector field $Z$ on $T^*M$ that satisfies $Z(H)=0$.
\end{def2}

An important class of symmetries which a Hamilton geometry admits are the symmetries associated to a constant of motion of the Hamilton dynamics, i.e. a quantity that is conserved along solutions of Hamilton equations $(x(\lambda),p(\lambda))$
\begin{equation}
	\frac{d}{d\lambda}S(x(\lambda),p(\lambda))=0\,.
\end{equation}
Noting that
\begin{equation}
	\frac{d}{d\lambda}S(x(\lambda),p(\lambda))=\dot x^a\partial_aS+\dot p_a\bar{\partial}^aS=\{S,H\}\,,
\end{equation}
we can equivalently say that a constant of motion is a function that Poisson commutes with the Hamiltonian.

The Poisson bracket of any two phase space functions $F$ and $G$ can be written in terms of a vector field associated to the functions
\begin{equation}
\{F,G\}=\partial_aF\bar{\partial}^aG-\bar{\partial}^aF\partial_a G=-(\bar{\partial}^aF\delta_a-\delta_a F\bar{\partial}^a)G=Z_F(G)=-Z_G(F)\,,
\end{equation}
So,  choosing the Hamiltonian $H$ as  one of the functions, we can write the symmetry condition as:
\begin{equation}\label{eq:psym}
Z_S(H)=(\bar{\partial}^aS\partial_a-\partial_aS\bar{\partial}^a)H=0\ \,. 
\end{equation}
Thus a constant of motion implies the infinitesimal diffeomorphism that is a symmetry of H
\begin{equation}
	\Phi_S(x,p)=(x^a+\epsilon\bar{\partial}^aS(x,p),p_a-\epsilon \partial_aS(x,p))\,.
\end{equation}
Note that this is a special case of (\ref{eq:diffeo}).

Another distinguished class of symmetries are the so-called manifold induced symmetries. A diffeomorphism of the base manifold $M$ can be represented infinitesimally by vector fields $X=\xi^a(x)\partial_a$ on $M$. It acts as a change of local coordinates $(x^a)\rightarrow(x^a+\xi^a)$. Such a local change of coordinates on $M$ induces a change of coordinates on $T^*M$ via $(x^a,p_a)\rightarrow(x^a+\xi^a,p_a-p_q\partial_a\xi^q)$, see equation (\ref{eq:mfcoord}). Thus a diffeomorphism on $M$ generated by the vector field $X$ induces a diffeomorphism on $T^*M$ generated by the vector field $X^C=\xi^a\partial_a-p_q\partial_a\xi^q\bar{\partial}^a$. In the literature $X^C$ is called the complete lift of $X$ from $M$ to $T^*M$.

\begin{def2}\textbf{Manifold symmetries}\\
	Let $(M,H)$ be a Hamilton geometry, $X=\xi^a(x)\partial_a$ be a vector field on the manifold and $X^C=\xi^a\partial_a-p_q\partial_a\xi^q\bar{\partial}^a$ be its complete lift to $T^*M$. A manifold symmetry of the Hamilton geometry is a diffeomorphism $\phi$ of $M$ whose generating vector field $X$ satisfies 
	\begin{equation}
	X^C(H)=0\,.
	\end{equation}
\end{def2}
\noindent Manifold symmetries are of particular interest in Hamilton geometry since they are the generalisation of the usual symmetries of a manifold in metric geometry. For the metric Hamiltonian $H_g$, defined in equation (\ref{eq:2}) the symmetry condition $X^C(H)=0$ becomes the condition that the Lie derivative  of $g$ with respect to $X$ has to vanish. To see this observe that for vector fields $X^C$ the symmetry condition (\ref{eq:symmcon}) can be rewritten in terms of~$g^H$
\begin{eqnarray}\label{eq:mfsymgh}
\frac{1}{2}\bar{\partial}^n \bar{\partial}^mX^C(H)&=&\frac{1}{2}(\xi^a\partial_a\bar{\partial}^n\bar{\partial}^mH-\partial_a\xi^n\bar{\partial}^m\bar{\partial}^aH-\partial_a\xi^m\bar{\partial}^n\bar{\partial}^aH-p_q \partial_a\xi^q\bar{\partial}^n\bar{\partial}^m\bar{\partial}^aH)\nonumber\\
&=&\xi^a\partial_ag^{Hmn}-\partial_a\xi^ng^{Hma}-\partial_a\xi^mg^{Hna}-p_q \partial_a\xi^q\bar{\partial}^ag^{Hmn}\,.
\end{eqnarray}
Now for $H=g^{ab}(x)p_ap_b$ the Hamilton metric satisfies $g^{Hab}(x,p)=g^{ab}(x)$, thus the  momentum derivative acting on the metric vanishes and we obtain in this case, due to the homogeneity of $H_g$, 
\begin{equation}
\frac{1}{2}p_np_m\bar{\partial}^n \bar{\partial}^mX^C(H)=p_mp_n\mathcal{L}_Xg^{mn}(x)=X^C(H)=0 \Leftrightarrow \mathcal{L}_Xg=0\,.
\end{equation}
Manifold induced symmetries always lead to a conserved phase space function. In virtue of equation (\ref{eq:psym}) the following holds
\begin{equation}
X^C(H)=0 \Leftrightarrow \{\xi^a(x)p_a,H\}=0\,.
\end{equation}
Examples of generators $X$ for manifold induced symmetries one may consider are the generators of spherical symmetry as they are used in Schwarzschild geometry or the cosmological symmetry generators which generate a homogeneous and isotropic geometry. We like to remark that for general symmetries $Z$ of the Hamilton geometry it is not possible to translate them into a condition on the Hamilton metric as it is possible for manifold symmetries in equation (\ref{eq:mfsymgh}). Thus symmetries in a Hamilton geometry are really characterised by the equation $Z(H)=0$ and not necessarily by conditions on the Hamilton metric. For the sake of overview we summarize the results of this symmetry section by displaying the three kinds of symmetries we distinguished:

\begin{itemize}
	\item A vector field on phase space generates a symmetry of $H$ if $Z(H)=0$
	\begin{equation}
		Z=\xi^a(x,p)\partial_a+\bar \xi_a(x,p)\bar{\partial}^a\,.
	\end{equation}
	\item Constants of motion $S(x,p)$ satisfy $\{S,H\}=0$ and induce symmetries
	\begin{equation}
	Z_S=\bar{\partial}^aS\partial_a-\partial_aS\bar{\partial}^a\,.
	\end{equation}
	\item Vector fields $X=\xi^a(x)\partial_a$ on spacetime generate symmetries of $H$ if $X^C(H)=0$
	\begin{equation}
	X^C=\xi^a\partial_a-p_q\partial_a\xi^q\bar{\partial}^a\,.
	\end{equation}
\end{itemize}

This section on symmetries of Hamilton spaces concludes our review and physical discussion of the Hamilton geometry of phase space. Next we study the geometry of the first order $q$-de Sitter dispersion relation and compare it to the geometry of the usual metric dispersion relation.

\section{Hamilton geometry of Planck-scale deformed dispersion relations - $q$-de Sitter-inspired example}\label{sec:moddr}
In this section we are going to show one explicit example of Hamiltonian geometry defined by a dispersion relation that describes a relativistic Planck-scale deformation of the propagation of particles on de Sitter spacetime. We will demonstrate explicitly how a change in the dispersion relation of freely falling point particles changes the phase space geometry and with it the geometry of spacetime. 

The model we consider is inspired by the $q$-de Sitter Hopf algebra \cite{Ballesteros:2013fh, Lukierski:1991ff, Ballesteros:2004eu, Lukierski:1991pn}, which is the only known fully consistent example of Planck-scale deformations of the de Sitter relativistic symmetries\footnote{In \cite{AmelinoCamelia:2012it} modifications of the de Sitter dispersion relation were also considered, but only at the single-particle level. The $q$-de Sitter Hopf algebra provides also a framework for describing particle interactions via the coproduct of the translation generators.}. This example is particularly interesting since the nontrivial interaction between spacetime curvature and Planck-scale effects are fully apparent. The quantum deformation parameter $q$ is a function of the two physical parameters entering the model: the expansion rate $h$ and the Planck-scale deformation parameter $\ell$ that is basically the inverse of the Planck energy ($\ell\sim 1/E_{P}$).  Different choices are possible for the actual dependence of $q$ on $h$ and $\ell$ \cite{Marciano:2010gq}: we will consider the  case where, in the limit where the spacetime curvature goes to zero, the model reduces to the much studied $\kappa$-Poincar\'e Hopf algebra \cite{Majid:1994cy, Lukierski:1992dt, Lukierski:1993wxa}, describing Planck-scale modifications of the special relativistic Poincar\'e group. The momentum space geometry encoded in the $\kappa$-Poincar\'e group has been shown to be the de Sitter one. Note that this statement is true when the momentum manifold is seen as the base manifold. Here, the momentum manifold is a part of the full phase space. Thus, while still finding that the momentum space associated to $\kappa$-Poincar\'e is curved, we should not expect it to have de Sitter curvature.
%

Before going to the actual example, we will first consider the general case of a metric Hamiltonian plus a Planck-scale perturbation term, and investigate the modifications of the phase space geometry to first order in the perturbation.

\subsection{First-order perturbation of metric Hamiltonian}\label{ssec:mdr3}
Consider the following modification of the metric Hamiltonian, governed by the parameter $\epsilon$, with dimensions of the inverse of an energy.
\begin{equation}
H=H_0+\epsilon H_1=g^{ab}(x)p_ap_b+\epsilon G^{abc}(x)p_ap_bp_c\,.
\end{equation} 
It induces via its level sets a modified dispersion relation. Introducing
\begin{equation}
g^{H_1 ab}=\frac{1}{2}\bar{\partial}^a\bar{\partial}^bH_1=3 G^{abc}p_c,\ g^{H_1}{}_{ab}=g_{ai}g_{bj}g^{H_1 ij}=3 G_{ab}{}^{c}p_c
\end{equation}
the Hamilton metric, see equation (\ref{eq:hammetric}), and its inverse can be calculated to first order in $\epsilon$ to be
\begin{equation}
g^{Hab}=g^{ab}+\epsilon g^{H_1 ab},\quad g^H_{ab}=g_{ab}-\epsilon g_{aj}g_{bi}g^{H_1 ij}=g_{ab}-\epsilon g^{H_1}_{ab}\,.
\end{equation}
With this notation the Hamilton non-linear connection coefficients, which we defined in equation (\ref{eq:nlin0}), become to first order in $\epsilon$
\begin{eqnarray}\label{eq:nlinex}
N_{ab}&=&-p_q\Gamma^q{}_{ab}+\epsilon \frac{3}{4}p_cp_d\big(g_{qb}\nabla_a G^{qcd}+g_{qa}\nabla_b G^{qcd}-2 g_{ma}g_{nb}g^{qc}\nabla_qG^{dmn}\big)\\
&=&-p_q\Gamma^q{}_{ab}+\epsilon p_q p_r T^{qr}{}_{ab}(x)\,,
\end{eqnarray}
where here $\nabla$ denotes the Levi-Civita covariant derivative of the metric $g_{ab}(x)$ and the last equality defines the tensor $T$ with components $T^{qr}{}_{ab}(x)$. Indeed we find that the non-linear connection coefficients are no longer linear in the momenta, as they are in metric phase space geometry  ($\epsilon=0$). The phase space curvature, see equation (\ref{eq:nlincurv}), which measures the integrability of spacetime as subspace of phase space can be calculated to be
\begin{equation}
R_{abc}(x,p)=p_qR^q{}_{abc}(x)+\epsilon p_qp_r(\nabla_c T^{qr}{}_{ab}(x)-\nabla_b T^{qr}{}_{ac}(x))\,.
\end{equation}
The zeroth order in $\epsilon$ is given by the Riemann tensor $R^q{}_{abc}(x)$ of the Levi-Civita covariant derivative of the metric $g$ and the first order correction, quadratic in the momenta, by the covariant derivatives of the tensor field $T$. The curvature of momentum space and spacetime as subsets of phase space, introduced in equations (\ref{eq:hcurv}) and (\ref{eq:vcurv}), defined by the associated connection coefficients~(\ref{eq:lincoef})
\begin{eqnarray}
\Gamma^{\delta a}{}_{bc}&=&\Gamma^a{}_{bc}(x)+\epsilon \frac{2}{3}p_q g^{ad}(\nabla_d G_{bc}{}^q-\nabla_b G_{cd}{}^q-\nabla_c G_{bd}{}^q)=\Gamma^a{}_{bc}+\epsilon p_q\gamma^{qa}{}_{bc},\\
C^{ab}{}_c&=&-\epsilon \frac{3}{2}G^{ab}{}_c\,,
\end{eqnarray}
become
\begin{eqnarray}\label{eq:curvex}
R^{Ha}{}_{bcd}(x,p)&=&R^a{}_{bcd}(x)+\epsilon p_q(\nabla_c\gamma^{qa}{}_{bd}-\nabla_d\gamma^{qa}{}_{bc}+\frac{3}{2}R^q{}_{rcd}G^{ra}{}_b)\\
Q_a{}^{bcd}(x,p)&=&0\,.
\end{eqnarray}
Observe that the spacetime curvature part is given by the Riemann tensor and is not dependent on the momenta in lowest order, while the momentum space curvature is at least of second order in~$\epsilon$. In fact, cubic terms in the momenta contribute to the curvature in the momentum space sector only at higher orders in $\epsilon$.

The phase space geometry we derived here is built from one function on phase space, the Hamiltonian, or equivalently from two tensors $g$ and $G$ on spacetime. Thus, at first order in the deformation parameter,  one can interpret the Hamilton geometry of phase space as multi-tensor geometry, from a spacetime point of view.

 As last part of this section we display the Hamilton equations of motion for the Hamiltonian above
\begin{equation}
\dot p_a+\partial_ag^{cd}(x)p_cp_d+\epsilon \partial_aG^{bcd}(x)p_bp_cp_d=0,\quad \dot x^a-2g^{ab}(x)p_b-3\epsilon G^{abc}(x)p_bp_c=0\,.
\end{equation}
To understand their relation to the autoparallel equation of the Hamiltonian phase space geometry to first order in $\epsilon$ we write the first one in the form (\ref{eq:hamauto})
\begin{eqnarray}
\dot p_a+(-p_m\Gamma^m{}_{ab}+\epsilon p_q p_r T^{qr}{}_{ab}(x))(2g^{bc}(x)p_c+\epsilon3G^{bcd}(x)p_cp_d)&=&-\delta_a H\\
\dot p_a -2 \Gamma^m{}_{ab}p_mg^{bc}p_c+\epsilon p_qp_rp_c(2T^{qr}{}_{ab}g^{bc}-3\Gamma^c{}_{ab}G^{bqr})&=&-\epsilon p_qp_rp_c (\nabla_a G^{qrc}-2 T^{rq}{}_{ab}g^{bc})\,.\nonumber
\end{eqnarray}
The left hand side of the last equation above is the term $\dot p_a+N_{ab}\bar{\partial}^bH$, which, set equal to zero, is the autoparallel equation of the geometry. Here we have a non-vanishing right hand side in the equation which prevents the test particles from moving along autoparallels of the geometry, due to the inhomogeneity of the Hamiltonian to first order in $\epsilon$. In zeroth order we recognize the geodesic equation of the metric Hamiltonian where the particles indeed propagate along autoparallels of the geometry.

After this general discussion of the modifications of phase space geometry induced by Planck-scale corrections to metric dispersion relations, we go on detailing the explicit example inspired from the $q$-de Sitter model.


\subsection{The $q$-de Sitter phase space geometry and its  $\kappa$-Poincar\'e limit} \label{ssec:qds}

The $q$-de Sitter Hopf algebra is characterised by a Casimir which can be encoded in the Hamiltonian \footnote{As mentioned before, we consider a specific realisation of the $q$-de Sitter model, where the quantum deformation parameter $q$ is fixed as a function of the two parameters $h$ and $\ell$ so that the two limits discussed in the following hold. See \cite{Marciano:2010gq} for details.}
\begin{equation}\label{eq:exH}
H_{qdS}(x,p)=H_0+\ell H_1=p_0^2-p_1^2(1+2 h x^0)-\ell p_0 p_1^2(1+2hx^0)\,,
\end{equation}
where we use $\ell$, the inverse of the Planck energy, as perturbation parameter in the momenta and $h$ is the expansion rate parameter. We work at first order in $\ell$ and $h$ and in $1+1$ dimensions. Note that $H_{0}$ here is the standard de Sitter Hamiltonian, written in flat slicing coordinates \cite{Amelino-Camelia:2013uya, Marciano:2010gq}:
\begin{equation}
H_{0}=p_0^2-p_1^2(1+2 h x^0)\,.
\label{eq:deSitterH}
\end{equation}
For $h=0$ the $q$-de Sitter Hamiltonian reduces to the $\kappa$-Poincar\'e Hamiltonian, written in the bicrossproduct basis \cite{Majid:1994cy}:
\begin{equation}
H_{\kappa P}=p_0^2-p_1^2-\ell p_0 p_1^2\,.
\label{eq:kPH}
\end{equation} 
Thus all results derived in this section immediately translate to the $\kappa$-Poincar\'e dispersion relation by setting $h=0$, or to the standard de Sitter dispersion relation for $\ell=0$.\footnote{The formal similarity between the geometries given by (\ref{eq:deSitterH}) and (\ref{eq:kPH}) have been explored in~\cite{Amelino-Camelia:2013uya}.}

The symmetries of the full Hamiltonian $H_{qdS}$ are induced by the following phase space functions which are constants of motion
\begin{eqnarray}
	P_0(x,p)&=&p_0 + h x^1 p_1,\ P_1(x,p)=p_1,\\
	 N(x,p)&=&p_1 x^0+p_0 x^1+ h \left(p_{1} (x^0)^2 + \frac{1}{2} p_1 (x^1)^2\right) - \ell \left(x^1 p_0{}^2 + \frac{1}{2}  x^1 p_1^2\right) \nonumber\\
	 &-& h \ell \left(p_1^2 x^0 x^1 + \frac{3}{2} p_0 p_{1} (x^1)^2 \right).
\end{eqnarray}
They all Poisson commute with the Hamiltonian (\ref{eq:exH}). Observe that $P_1$ is also a manifold induced symmetry according to our definition in section \ref{ssec:symm}, while $P_0$ and $N$ are not.  Of course these symmetries are the same provided within the $q$-de Sitter Hopf algebra framework.

We now derive the building blocks of the phase space geometry for this Hamiltonian. The $H$-metric can easily be obtained as
\begin{eqnarray}
g^{Hab}=\frac{1}{2}\bar{\partial}^a\bar{\partial}^bH_{qdS} &=&\left(\begin{array}{lll}1 &\;\;& -\ell p_1 (1+2 h x^0) \\
-\ell p_1 (1+2 h x^0) &\;\;&- (1+2 h x^0)(1+ \ell p_0) \\ \end{array}\right)\,,\\
g^{H}{}_{ab}& =&\left(\begin{array}{lll}1 &\;\;& -\ell p_1 \\ -\ell p_1 &\;\;& -(1-2 h x^0)(1- \ell p_0)\end{array}\right)\,.
\end{eqnarray}
The non-linear connection coefficients, directly calculated from their defining equation (\ref{eq:nlin0}) are
\begin{equation}
N_{ab}=\left(
\begin{array}{cc}
h  \ell p_1^2& h p_1 \\
h p_1 & h p_0 (1-\ell p_{0})\\
\end{array}
\right)\,.
\end{equation}
Observe that they vanish in the $\kappa$-Poincar\'e limit, since there $h=0$. The same is true for the horizontal connection coefficients $\Gamma^{\delta a}{}_{bc}$ (\ref{eq:lincoef}). In the $q$-de Sitter phase space they are 
\begin{eqnarray}
\Gamma^{\delta0}{}_{00}&=&0,\quad \Gamma^{\delta0}{}_{01}=-  h \ell p_{1},\quad \Gamma^{\delta0}{}_{11}=-h(1-2\ell p_{0}) \\
\Gamma^{\delta1}{}_{00}&=&- h \ell p_1,\quad \Gamma^{\delta1}{}_{01}=-h ,\quad\Gamma^{\delta1}{}_{11}=\frac{3}{2}  h \ell p_{1}\,.
\end{eqnarray}
And thus the curvature of spacetime as subspace of phase space contains a momentum dependent part at first order in $\ell$ as we anticipated in equation (\ref{eq:curvex}). In the $\kappa$-Poincar\'e limit the spacetime curvature vanishes, so the spacetime is flat. The coefficients for the momentum space curvature are derived from $-\frac{1}{2}\bar\partial^a g^{H cb}=C^{abc}$ and neither vanish in the $q$-de Sitter case nor its  $\kappa$-Poincar\'e limit:
\begin{equation}
C^{000}=0,\quad C^{001}=0,\quad C^{011}=\frac{\ell}{2}(1+2 h x^0),\quad C^{111}=0
\end{equation} 
and thus $-\frac{1}{2}g^H_{rc}\bar\partial^a g^{H rb}=C^{ab}{}_c$
\begin{eqnarray}\label{eq:Cel1}
C^{00}{}_0&=&0,\quad C^{00}{}_1=0,\quad C^{01}{}_0=0,\quad C^{11}{}_1=0\,,\\
C^{01}{}_1&=&-\frac{\ell}{2},\quad C^{11}{}_0=\frac{\ell}{2}(1+2 h x^0)\,.\label{eq:Cel2}
\end{eqnarray} 
The momentum space curvature is, as explained in the previous section in equation (\ref{eq:curvex}), of order~$\ell^2$. 

The Hamilton equations of motion (\ref{eq:heom}) read:\footnote{A complete analysis of the kinematics implied by these equations of motion will appear in \cite{qdSpheno}.}
\begin{eqnarray}
\dot x^{0}-2 p_{0}+\ell p_{1}^{2}(1+2 h x^{0})&=&0 \, ,\\
\dot x^{1}+2 p_{1}(1+2 h x^{0})+2\ell p_{0}p_{1}(1+2 h x^{0})&=&0 \, ,\\
\dot p_0- 2 h p_1^2-2 h \ell p_0 p_1^2 &=&0\\
\dot p_1&=&0 \,.
\end{eqnarray}
As in the previous section we write the last two ones in the form (\ref{eq:hamauto}):
\begin{eqnarray}
\dot p_0- 2 h p_1^2&=&2 h \ell p_0 p_1^2\label{eq:hamautoex1}\\
\dot p_1-\ell h p_1^3 &=&-h \ell  p_1^3 \,.\label{eq:hamautoex2}
\end{eqnarray}

Due to our previous studies it is no surprise that they are not autoparallels, what they would be if the left hand side of the above equations were zero. There is a force-like term which drags the particles away from autoparallel motion due to the different homogeneities of the different terms $H_0$ and $H_1$ in the Hamiltonian. 

\section{An overview on the possible paths to quantization}\label{sec:quantisation}

At this point, some remarks might be in order, concerning the manifestly classical nature
of our framework. As mentioned already in the introduction, we are here concerned with a regime where $\hbar\rightarrow 0$ in such a way that the Planck energy stays finite. 
As it stands, the considered deformations of the classical dispersion relations can be viewed as 
capturing in an effective way some important features of a more complete theory of quantum
gravity. However, for applications to Planck scale phenomenology it is 
important to see explicitly how deep into the quantum regime Hamilton geometry carries. When $\hbar\neq 0$, for both the $\kappa$-Poincer\'e and $q$-de Sitter models there exists an associated homogeneous spacetime (for the $\kappa$-Poincer\'e case this is known as $\kappa$-Minkowski space \cite{AmelinoCamelia:2012ra, AmelinoCamelia:2011nt}). Most strikingly, in this case one finds non-vanishing commutation relations between
functions on spacetime---a feature that still awaits rigorous operational interpretation \cite{AmelinoCamelia:2012ra}, especially
in a generally covariant approach as ours, where coordinates should lose any meaning that 
goes beyond that of being mere labels of events. The noncommutativity of the functions of
spacetime coordinates is usually encoded in modified
commutation relations between coordinates operators. This requires a deformed symplectic structures
to properly enforce the Jacobi identities. As pointed out before (see sec.~\ref{ssec:21hs}), at the classical
level it is always possible to map a non trivial symplectic structure into the canonical one,
the hamiltonian being mapped accordingly~\cite{Nozari:2014qja,Amelino-Camelia:2013uya,AmelinoCamelia:2011cv}.
At this level the kinematics is equivalent and the symplectic structure is only a matter of choice.
And yet, one would hope that in the context of 
Hamilton geometry a generalised procedure exists\footnote{In order to realise such a scenario a possible strategy
would identify vector fields on $T^{*}M$ which are invariant under the symmetries
of the Hamiltonian and induce translations along the fibres, i.e. translate in momentum space.
Those could be regarded as the natural candidates for position operators in a quantum 
theory. A more careful analysis in this direction is left to the future.}, which in small spacetime regions is able to 
reproduce features of $\kappa$-Minkowski space. Having this procedure at hand, one could try to apply
a quantization procedure used in the flat spacetime case.\\
Literature offers quite a few examples of possible quantized versions of such flat
spacetime limit. These can be a source of inspiration. There are two
main attempts to a quantized theory: one aims to obtain a relativistic quantum mechanics of the
classical theory; another one aims at a (possibly quantum) field theory on a noncommutative bundle.
The first~\cite{AmelinoCamelia:2012ra} attempt points towards the construction of a proper set of states able to represent freely
moving particles, enforcing a $\kappa$-Poincar\'e version of the relativistic symmetries.\\
The other approach~\cite{AmelinoCamelia:2002mu} tries to give a path integral description of a field 
theory over a noncommutative spacetime. More in general, the aim is to get the quantum dynamics of a
field over a phase space whose symmetries are deformed with respect to the special relativistic ones.\\
Since the same classical model can be obtained as classical limit of many different quantized frameworks, one should be very cautious in choosing one or another approach to quantization, without a serious scrutiny of the physics that those different frameworks would imply.

\section{Conclusions and outlook}\label{sec:4conc}
With the interpretation of a dispersion relation as level set of a Hamilton function on phase space we were able to derive the Hamilton geometry of phase space directly from the dispersion relation. We identified the spacetime and the momentum space as subspaces of phase space and consistently described their geometry. It turned out that for a general dispersion relation, i.e. a general Hamiltonian, the geometric objects of spacetime and momentum space, like covariant derivatives and curvature, depend on positions and momenta. That means that in general it is not possible to disentangle the geometry of spacetime and the geometry of momentum space: they are intertwined as parts of the geometry of phase space. A disentanglement of the geometries into a geometry of spacetime that is not momentum depend and a geometry of momentum space that is not position dependent, is only possible for very special dispersion relations: for Hamiltonians whose third derivative and higher order derivatives with respect to the momenta  vanish the momentum space geometry is flat and the spacetime geometry is the usual metric spacetime geometry, while for Hamiltonians that do not depend on the spacetime coordinates we find a flat spacetime and a possibly curved momentum space. 

We have observed that  Hamilton geometry  can be effective in the description of the phase space geometry when Planck-scale modifications of particles' dispersion relations are introduced. This is especially useful when modifications are introduced for particles moving in a curved spacetime, which is a relevant case for phenomenological purposes, but until now has been very difficult to describe in a coherent framework.

The Hamilton equations of motion, which determine the trajectories of test particles through phase space, become the autoparallel equation of the phase space geometry with a source term, so that the effect of Planck-scale modifications on phase space geometry is to drag particles away from purely geometric free fall. This is characteristic of Planck-scale modifications: in theorem \ref{thm:2} we found that for all Hamiltonians that are homogeneous with respect to the momenta these source terms vanish and the test particles propagate on autoparallels through phase space.

In order to give explicit examples of how phase space geometry is modified at the Planck-scale, we analysed, at the first order in the perturbation parameters,  the phase space geometry associated to the dispersion relation of $q$-de Sitter Hopf algebra, which includes as limit the phase space geometry of the $\kappa$-Poincar\'e Hopf algebra dispersion relation. The $q$-de Sitter phase space geometry yields a curved spacetime manifold and a curved momentum space. We showed explicitly that spacetime curvature depends on both positions and momenta. The momentum space curvature is non-zero, but of second order in the Planck-scale deformation parameter $\ell$, so we did not give an explicit expression. In the $\kappa$-Poincar\'e limit of the $q$-de Sitter Hamiltonian  the spacetime becomes flat and the momentum space is curved. Since the $q$-de Sitter Hamiltonian is inhomogeneous in the momenta the Hamilton equations of motions are not the autoparallels of the geometry but contain a force-like term. 
Symmetries in the Hamilton geometry framework are simply the symmetries of the Hamiltonian, and we have shown that in the $q$-de Sitter case they are the same as the ones described by the Hopf algebra, at the single particle level.

We have not yet developed a coherent description of particles' interactions. 
This is expected to be non-trivial, since we know that relativistic 
compatibility would require to modify energy-momentum conservation. We 
learned from the Hamilton geometry framework that if both spacetime and 
momentum space are non-trivially curved we have to consider tensor fields on 
phase space to describe the motion of particles. The challenge is then to 
identify the appropriate tensors on phase space which describe the interaction 
of point particles. We seek for an appropriate representation of the momentum 
of a particle, classically being a one form on spacetime, and now to be 
generalized in terms of a tensor field on phase space. Having clarified such an 
identification we will investigate how the interaction of particles can be 
formulated through these phase space tensors. We imagine that a realisation of 
the addition of momenta in Hamilton phase space geometry could be possible by 
parallel transport of the momenta, identified as tensors on phase space, along 
autoparallels in momentum space, similarly to what is done in the relative 
locality context, where momentum space is taken as curved base manifold and 
spacetime is flat \cite{AmelinoCamelia:2011bm, Amelino-Camelia:2013sba}. Based 
on the foundations laid in this article we plan to address this issue soon in a 
further publication.

Another feature that has to be clarified when dealing with a non-metric phase space geometry is the description of observers. The observer frames used in general relativity clearly have to be modified since in general there is no underlying spacetime metric to obtain such frames. It is necessary to define observers directly from the Hamiltonian instead. One approach to do so is to use the notion of radar orthogonality to construct a space-time split for each observer worldline, a procedure that has been successfully applied to Finsler spacetimes in \cite{Pfeifer:2014yua}. Another interesting approach based on the analysis of the geometry of observer space in terms of Cartan geometry is proposed in \cite{Gielen:2012fz, Gielen:2012pn, Hohmann:2013fca}.

Apart from quantum gravity phenomenology general Hamiltonians appear in the geometric optics limit of the study of partial differential equations. They determine the propagation of ray solutions of the partial differential equation. Here the Hamilton geometry leads directly to a geometric understanding of the trajectories along which the ray solutions propagate since the Hamiltonians are homogeneous and thus the rays propagate along autoparallels of the phase space geometry. 

\acknowledgments
CP gratefully thanks the Center of Applied Space Technology and Microgravity (ZARM) at the University of Bremen for hospitality and support. GG, LB and NL are supported by grants from the John Templeton Foundation. LKB acknowledges the support by a Ph.D. grant of the German Research Foundation within its Research Training Group 1620 \emph{Models of Gravity} and thanks the Niels Bohr Institute at Copenhagen for its hospitality.

\appendix

\section{Proof of theorem 2}\label{app:prfthm2}
In section \ref{ssec:hconn} we introduced the Hamilton non-linear connection and discussed its properties. For homogeneous connections we claimed in Theorem \ref{thm:2} that
\begin{eqnarray}
	\delta_a H=\partial_a H-N_{ab}\bar{\partial}^bH=0\,.
\end{eqnarray}
Here we display the proof of the Theorem.
\begin{prf}
	For homogeneous functions Euler's Theorem holds, see for example \cite{BCS} for a proof, 
	\begin{equation}
	p_a\bar\partial^a H=r H\,,
	\end{equation}
	and thus yields the following relations
	\begin{equation}
	\bar\partial^a H= g^{H aq}p_q\frac{2}{r-1},\quad H=\frac{2}{r(r-1)}g^{Haq}p_qp_a\,.
	\end{equation}
	Using these we derive
	\begin{eqnarray}
	N_{ab}\bar\partial^b H&=& \frac{1}{4}\bigg(\bar\partial^b H\bar\partial^iH\partial_i g^H_{ab}-\frac{2}{r-1}p_c g^{Hcb}\partial_iH\bar\partial^i g^H_{ab}+\bar\partial^b Hg^H_{ai}\partial_b\bar\partial^i H+\frac{2r}{r-1}\partial_aH\bigg)\nonumber\\
	&=&\frac{1}{4}\bigg(-\partial_i\bar\partial^b H\bar\partial^iH g^H_{ab}+\frac{2}{r-1}p_c\partial_iH g^H_{ab}\bar\partial^ig^{H cb}+\bar\partial^b Hg^H_{ai}\partial_b\bar\partial^i H+\frac{2r}{r-1}\partial_aH\bigg)\nonumber\\
	&=&\frac{1}{4}\bigg(\frac{2(r-2)}{r-1}\partial_aH+\frac{2r}{r-1}\partial_aH\bigg)=\partial_aH\,.\ \square
	\end{eqnarray}
\end{prf}

\section{The geometry of the cotangent bundle}\label{app:cotb}
The mathematical language we used to derive the geometry of phase space from a Hamiltonian, respectively a dispersion relation, in section \ref{sec:2HamGeom} is the geometry of the cotangent bundle. Since we aimed to not overload the main text of this article with technical mathematics we add some details on the general geometry of the cotangent bundle here. To make this appendix optimally and self contained readable there may appear some repetitions from the main text. In particular we emphasize the role of the bundle structure of the cotangent bundle i.e. the local split into fibres and base manifold, physically speaking into momentum and position space. The building block of the geometry is a general connection that splits the tangent and cotangent spaces of the bundle into horizontal and vertical parts. Again, in physical words, into tangent spaces along spacetime and tangent space along momentum space. This then leads to the notion of curvature and covariant derivatives. The mathematical concepts we present here are a particular application of general connections on fibre bundle, see for example \cite{NODG}, to the cotangent bundle.

\subsection{The cotangent bundle in manifold induced coordinates}\label{app:cotbcoord}
As we mentioned in the main text in section \ref{ssec:21hs} the cotangent bundle of an $n$-dimensional manifold $M$ is itself a $2n$-dimensional manifold $T^*M$. It is the union of all cotangent spaces of $M$
\begin{equation}
T^*M=\bigcup_{q\in M}T^*_qM\,.
\end{equation}
It carries the natural structure of a fibre bundle with total space $T^*M$, fibre $\mathbb{R}^n$ and projection map $\pi$ that associates to each one-form $\Omega\in T^*_qM$ the point $q\in M$. In a local coordinate chart $(U,x)$ around $q\in M$ we can expand the one-form $\Omega$ in these coordinates as $\Omega=p_a dx^a_{|x}$. The components $p_a$ of $\Omega$ with respect to the local coordinate basis of $T_qM$ and the coordinates $x^a$ of the base point $q\in M$ can now be used as so-called locally manifold induced coordinates of $T^*M$ around $T^*_qM\subset T^*M$. In these coordinates we write $\Omega=p_a dx^a_{|x}=(x,p)\in T^*M$. From now on we consider the cotangent bundle in manifold induced coordinates, exceptions are stated explicitly. Changing the coordinates on the base manifold $M$ from $x$ to $\tilde x(x)$ induces a coordinate change of the manifold induced coordinates on the cotangent bundle according to the transformations of one-form components on the manifold $\Omega=p_a dx^a_{|x}=p_a \frac{\partial x^a}{\partial \tilde x^b}(x(\tilde x))d\tilde x^b_{|x(\tilde x)}=\tilde p_b d\tilde x^b_{|\tilde x}$
\begin{equation}\label{eq:mfcoord2}
(x,p_a)\rightarrow (\tilde x(x,p),\tilde p_a(x, p))=\bigg(\tilde x(x),p_b \frac{\partial x^b}{\partial \tilde x^a}(x)\bigg)\,.
\end{equation}
The manifold induced coordinates lead directly to the coordinate basis of the tangent $T_{(x,p)}T^*M$ and cotangent $T^*_{(x,p)}T^*M$ spaces of the cotangent bundle. They are spanned by
\begin{equation}
T_{(x,p)}T^*M=\bigg\langle \frac{\partial}{\partial x^a}, \frac{\partial}{\partial p_b}\bigg\rangle=\langle \partial_a, \bar\partial^b \rangle,\quad T^*_{(x,p)}T^*M=\langle  dx^a, dp_b\rangle \,,
\end{equation}
where we introduced the short hand notations $\partial_a$ for the part of the coordinate basis of $T_{(x,p)}T^*M$ corresponding to the coordinates $x^a$ of the base manifold and $\bar\partial^a$ for the part of the coordinate basis of $T_{(x,p)}T^*M$ corresponding to the coordinates $p_a$. Their transformation behaviour under a change of the manifold induced coordinates yields
\begin{eqnarray}
( \partial_a, \bar\partial^a ) &\rightarrow& ( \tilde \partial_a, \tilde{\bar\partial}^a )=( \tilde\partial_a x^b \partial_b + \tilde \partial _a p_b \bar\partial^b, \partial_b \tilde x^a \bar\partial^b)\\
( dx^a, dp_a) &\rightarrow& ( d\tilde x^a, d\tilde p_a)=( \partial_b \tilde x^a dx^b, \partial _b \tilde p_a dx^b+\bar\partial^b\tilde p_a dp_b)
\end{eqnarray}
Due to the fact that the coordinate transformation is induced by the coordinate transformation matrix $\tilde\partial_a x^b$ on the base manifold which only depends on the $x$ respectively $\tilde x$ coordinates we see that the $\bar\partial^a$ and $dx^a$ transform just like tensors on the base manifold, while the $\partial_a$ and the $dp_a$ transform in a more complicated way. The $\bar\partial^a$ span the tangent space to the fibres of $T^*M$ which happens to be the kernel of the differential $d\pi$ of the fibre bundled projection $\pi$. In fibre bundle language one calls this set the vertical tangent space $V_{(x,p)}T^*M$ of the cotangent bundle. It is annihilated by the so-called horizontal cotangent space $H^*_{(x,p)}T^*M$ of the cotangent bundle which is spanned by the $dx^a$ part of the basis of $T^*_{(x,p)}T^*M$, i.e. $dx^a(\bar\partial^b)=0$. It is possible to obtain a complete basis of $T_{(x,p)}T^*M$ and $T^*_{(x,p)}T^*M$ which transforms under manifold induced coordinate transformations like tensors on the base manifold by the introduction of a connection on $T^*M$. Such a connection defines the geometry of the bundle which respects the bundle structure.

\subsection{Connections on the cotangent bundle, curvature and autoparallels}\label{app:cotbconn}
In section \ref{ssec:hconn} we introduced a distinguished unique connection on the cotangent bundle, which enabled us to study the geometry of momentum and position space as subspaces of phase space.

A connection on the cotangent bundle is a projection from the complete tangent space of the cotangent bundle onto the vertical tangent space, which we just introduced above.
\begin{def2}
	A connection one-form $\omega$ on the cotangent bundle is a projection, $\omega \circ \omega = \omega$,
	\begin{equation}
	\omega_{(x,p)}: T_{(x,p)}T^*M \rightarrow V_{(x,p)}T^*M\,.
	\end{equation}
	Expressed in manifold induced coordinates it takes the form
	\begin{equation}
	\omega_{(x,p)}=(dp_a+N_{ab}(x,p)dx^b)\otimes \bar\partial^a\,.
	\end{equation}
	The components $N_{ab}(x,p)$ are called connection coefficients of $\omega$.
\end{def2}
\noindent A general connection $\omega$ is also-called non-linear connection to clarify the difference to affine connection geometry. This means for a general connection the connection coefficients $N_{ab}(x,p)$ may depend quite arbitrarily on the fibre coordinates $p_a$ where in affine connection geometry they are linear in the $p_a$ and can be written as $N_{ab}(x,p)=p_c\Gamma^c{}_{ab}(x)$. Then the $\Gamma^a{}_{bc}(x)$ are the connection coefficients of an affine connection, for example they may be the Christoffel symbols of the Levi-Civita connection used in metric geometry. We calculate the change of the connection coefficients under a manifold induced coordinate transformation by comparing $\omega=(dp_a+N_{ab}(x,p)dx^b)\otimes \bar\partial^a=(d\tilde p_a+\tilde N_{ab}(\tilde x,\tilde p)d\tilde x^b)\otimes\tilde{ \bar\partial}^a$. We find
\begin{equation}
\tilde N_{cm}=\tilde\partial_c x^q\tilde \partial_m p_q+N_{ab}\tilde\partial_m x^b \tilde \partial_c x^a\,.
\end{equation}
This transformation behaviour of the connection coefficients enable us to find the complete basis of $T_{(x,p)}T^*M$ and $T^*_{(x,p)}T^*M$ which transforms like tensor components on the base manifold under manifold induced coordinate changes. Introducing the linear combinations of the coordinate bases 
\begin{equation}\label{eq:horrder}
\delta_b=\partial_b- N_{ab}\bar\partial^a,\ \delta p_b=dp_b+N_{ba}dx^a\,,
\end{equation}
where the $\delta_a$ span the kernel of the connection one-form $\omega$ and the $\delta p_a$ annihilate this kernel $\delta p_a(\delta_b)=0$, we find the new complete basis with desired transformation behaviour $\delta_a=\partial_a \tilde x^b \tilde\delta_b$ and $\delta p_a=\tilde\partial_m x^b \delta p_b$
\begin{equation}
T_{(x,p)}T^*M=\langle \delta_a, \bar\partial^a \rangle,\ T^*_{(x,p)}T^*M = \langle dx^a, \delta p_a \rangle \,.
\end{equation}
In standard fibre bundle language the span of the $\delta_a$ is called horizontal tangent space $H_{(x,p)}T^*M$ of the cotangent bundle and the span of the $\delta p_a$ the vertical cotangent space $V_{(x,p)}^*T^*M$ of the cotangent bundle. Thus a connection on the cotangent bundle enables us to split the tangent and cotangent spaces of the cotangent bundle in vertical and horizontal part. The vertical space represents the tangent respectively cotangent space of the fibre and the horizontal part represents the tangent respectively cotangent spaces of the base manifold in the tangent respectively cotangent spaces of the cotangent bundle.

Recall that an $(n,m)$ $d$-tensor field $T$ is a tensor field on the cotangent bundle for which the following holds
\begin{equation}
T(X_1,...,X_m, \Lambda_1,...,\Lambda_n)=T(P_1(X_1),...,P_m(X_m), P^1(\Lambda_1),...,P^n(\Lambda_n))\,,
\end{equation}
where $P_i$ is a projector on the horizontal or vertical tangent bundle of the cotangent bundle and $P^i$ is the projector on the horizontal or vertical cotangent bundle of the cotangent bundle. For example the components of the Hamiltonian metric, defined in equation (\ref{eq:hammetric}), define $(0,2)$ $d$-tensor fields like $g^{Hab}(x,p)\delta p_a\otimes \delta p_b$ or $g^H_{ab}(x,p)dx^a\otimes dx^b$. To find the desired unique non-linear connection we introduce a so-called dynamical covariant derivative $\nabla$ that acts on $d$-tensor fields $T$ with components $T^{a_1...a_n}{}_{b_1...b_m}(x,p)$ as follows
\begin{eqnarray}\label{eq:dcov}
\nabla T^{a_1...a_n}{}_{b_1...b_m}&=&(\bar\partial^qH\partial_q-\partial_pH\bar\partial^p)T^{a_1...a_n}{}_{b_1...b_m}+Q^{a_1}{}_cT^{c...a_n}{}_{b_1...b_m}+...+Q^{a_n}{}_cT^{a_1...c}{}_{b_1...b_r}\nonumber\\
&-&Q^{c}{}_{b_1}T^{a_1...a_n}{}_{c...b_m}-...-Q^{c}{}_{b_m}T^{a_1...a_n}{}_{b_1...c}\,.
\end{eqnarray}
Observe that the differential operator acting on the components of the $d$-tensor field is given by the Poisson bracket between $H$ and the components. Further details on the dynamical covariant derivative, which is usually introduced on the tangent bundle of a manifold, can be found in the book by Bucutaru and Miron \cite{FLG}. Here we used the Legendre transform to define a dynamical covariant derivative directly on the cotangent bundle. The $Q^a{}_b$ are the connection coefficients of a non-linear connection on the tangent bundle that is dual to the non-linear connection we seek to find on the cotangent bundle
\begin{equation}
Q^a{}_b=2 N_{qb}g^{H qa}-\bar\partial^a\partial_bH\,.
\end{equation}
The derivation of the $Q^a{}_b$ can be found in the appendix \ref{app:dualconn}.

The curvature of a connection, introduced for the Hamilton connection in equation (\ref{eq:nlincurv}), is a measure of integrability of the horizontal bundle, that is the union of all horizontal tangent spaces. 

\begin{def2}\textbf{The curvature of the Hamilton non-linear connection}\label{def:curvature}\\
	Let $(M,H)$ be a Hamilton geometry. We call
	\begin{equation}
	[\delta_a,\delta_b]=\big(-\delta_a N_{cb}+\delta_b N_{ca}\big)\bar\partial^c=R_{cab}\bar\partial^c
	\end{equation}
	the curvature of the connection. 
\end{def2}
\noindent By Frobenius' theorem, see for example \cite{Wald}, this object indeed measures the integrability of the horizontal bundle. It is integrable if and only if $R_{abc}=0$. This means the spacetime manifold $M$ is a submanifold of phase space $T^*M$ if and only if the curvature of the connection $\omega$ vanishes. The curvature of the non-linear connection does not require further structures and is completely determined by the non-linear connection. 

Autoparallels of the connection are curves $\gamma:\mathbb{R}\rightarrow T^*M$ with purely horizontal tangent. Thus $\gamma$ is an autoparallel if it satisfies 
\begin{equation}\label{eq:autopara}
\dot p_a+N_{ab}(x,p)\dot x^b=0\,,
\end{equation}
since then its tangent is indeed purely horizontal
\begin{equation}
\gamma(t)=(x(t), p(t)) \Rightarrow \dot\gamma= \dot x^a\partial_a+\dot p_a \bar\partial^a=\dot x^a\delta_a+(\dot p_a+N_{ab}\dot x^b)\bar\partial^a= \dot x^a\delta_a\,.
\end{equation}
As final remark of this section we want to mention that a connection is called compatible with the canonical symplectic structure of the cotangent bundle if it is symmetric, i.e. if $N_{ab}=N_{ba}$. This condition ensures that the canonical symplectic form vanishes on $H_{(x,p)}T^*M$. To see this we write the canonical symplectic form $\Omega$ in the horizontal-vertical basis of $T_{(x,y)}T^*M$
\begin{equation}\label{eq:sympl}
\Omega=dp_a\wedge dx^a=\delta p_a \wedge dx^a - N_{ab} dx^b \wedge dx^a\,.
\end{equation}
Since $\delta p_a(\delta _b)=0$ we have that $\Omega(\delta_a, \delta_b)=0$ if and only if $N_{ab}-N_{ba}=0$. This is the cotangent bundle version of the torsion-freeness condition that is employed in metric geometry and we used to find the Hamilton non-linear connection in definition \ref{def:nonlin} and theorem \ref{thm:1}.

\section{Dual connections}\label{app:dualconn}
In appendix \ref{app:cotbconn} we discussed that a connection $\omega$ on the cotangent bundle induces a split of the tangent spaces of the cotangent bundle into horizontal and vertical subspace
\begin{equation}
T_{(x,p)}T^*M=H_{(x,p)}T^*M\oplus V_{(x,p)}T^*M=<\delta_a>\oplus<\bar\partial^a>\,.
\end{equation}
In Hamilton geometry we introduced the duality map $\sharp$ in equation (\ref{eq:dual}) which maps the cotangent bundle of the manifold to the tangent bundle. A connection $\omega$ on the cotangent bundle defined through connection coefficients $N_{ab}(x,p)$ is called dual to a connection $\omega'$ on the tangent bundle defined through connection coefficients $N^a{}_b(x,y)$ if the duality map maps the horizontal tangent spaces of the cotangent bundle onto the horizontal tangent spaces of the tangent bundle. Here $(x,y)$ denote the manifold induced coordinates of the tangent bundle \cite{Miron}. The horizontal tangent spaces of the cotangent and the tangent  bundle are spanned respectively by 
\begin{equation}
\delta_a=\partial_a-N_{ab}(x,p)\bar{\partial}^b\,,\quad \delta'_a=\partial_a-N^b{}_a(x,y)\bar{\partial}_b\,,
\end{equation}
where $\bar{\partial}_a=\partial/\partial y^a$. Thus the condition that two connections are dual is
\begin{equation}
d\sharp_{(x,p)}(\delta_a)=\delta'_a{}_{|\sharp(x,p)}\,.
\end{equation}
Recall the definition of $\sharp$ from which we derive the action of its differential
\begin{eqnarray}
\sharp:T^*M&\rightarrow& TM\nonumber\\
(x,p)&\mapsto& \sharp(x,p)=(x,\bar{\partial}^aH(x,p))=(x,y^a(x,p))\\
d\sharp_{(x,p)}:T_{(x,p)}T^*M&\rightarrow& T_{\sharp(x,p)}TM\nonumber\\
Z=Z^a\partial_a+\bar Z_a \bar{\partial}^a&\mapsto& d\sharp_{(x,p)}(Z)\\
&=&Z^ad\sharp_{(x,p)}(\partial_a)+\bar Z_ad\sharp_{(x,p)}(\bar{\partial}^a)\\
&=&Z^a(\partial_a+\partial_a\bar{\partial}^qH\bar{\partial}_q)+\bar Z_a \bar{\partial}^a\bar{\partial}^qH \bar{\partial}_q\,.
\end{eqnarray}
Applying this mapping to the horizontal basis vectors in $H_{(x,p)}T^*M$ yields
\begin{eqnarray}
d\sharp_{(x,p)}(\delta_a)&=&d\sharp_{(x,p)}(\partial_a)-d\sharp_{(x,p)}(N_{ab}(x,p)\bar{\partial}^b)=\partial_a+\partial_a\bar{\partial}^qH\bar{\partial}_q-N_{ab}\bar{\partial^b}\bar{\partial}^qH\bar{\partial}_q\nonumber\\
&=&\partial_a-(2 N_{ab}g^{Hbq}-\partial_a\bar{\partial}^qH)\bar\partial_q\,.
\end{eqnarray}
Thus in order to have a dual connection on the tangent bundle the connection coefficients have to be related to the connection coefficients of the connection on the cotangent bundle by $N^a{}_b=2 N_{bq}g^{Hqa}-\partial_b\bar{\partial}^aH$. These are, as claimed, exactly the $Q^a{}_b$ which we introduced in section \ref{ssec:hconn} when we defined the dynamical covariant derivative in equation \ref{eq:dyncov} and in the previous appendix~\ref{app:cotbconn}.

\bibliographystyle{utphys}
\bibliography{HG-1}

\end{document}